\documentclass{aadebug} 
\usepackage{theorem} 
\usepackage{programlist} 
\usepackage{amsmath}

\corr{0309027}{55}

\newtheorem{definition}{Definition}

\newcommand{\lfp}{\mathop{\mathrm{lfp}}\nolimits}

\newcommand{\stmt}[1]{{\small\fbox{#1}}}
\newcommand{\ab}[1]{\ensuremath{ab(\stmt{#1})}}
\newcommand{\nab}[1]{\ensuremath{\neg\ab{#1}}}
\newcommand{\syn}[2]{\ensuremath{syn(\stmt{#1},#2)}}
\newcommand{\nosyn}[1]{\ensuremath{\neg syn(\stmt{#1})}}

\begin{document}
\pdfinfo{
/Title (Model-Based Debugging using Multiple Abstract Models)
/Author (Wolfgang Mayer and Markus Stumptner)
}

\runningheads{Wolfgang Mayer et al.}{Model--Based Debugging using
  Multiple Abstract Models}

\title{Model--Based Debugging using Multiple Abstract Models}

\author{Wolfgang Mayer\addressnum{1}, Markus Stumptner\addressnum{1}\,\extranum{1}}

\address{1}{
  University of South Australia\\ 
  Advanced Computing Research Centre\\ 
  Mawson Lakes, Adelaide SA 5095, Australia}

\extra{1}{E-mail:~\{mayer,mst\}@cs.unisa.edu.au}

\begin{abstract}
  This paper introduces an automatic debugging framework that relies
  on model--based reasoning techniques to locate faults in programs.
  In particular, model--based diagnosis, together with an abstract
  interpretation based conflict detection mechanism is used to derive
  diagnoses, which correspond to possible faults in programs. Design
  information and partial specifications are applied to guide a model
  revision process, which allows for automatic detection and
  correction of structural faults.
\end{abstract}


\section{Introduction}

Detecting a faulty behavior within a program, locating the cause of
the fault, and fixing the fault by means of changing the program,
continues to be a crucial and challenging task in software
development. Many papers have been published so far in the domain of
detecting faults in software, e.g., testing or formal
verification~\cite{cdhprlz00}, and locating them, e.g., program
slicing~\cite{wei84} and automatic program debugging~\cite{lloyd87b}.
More recently model--based diagnosis~\cite{rei87} has been used for
locating faults in software~\cite{con93,msww02a}.

This paper extends previous research in several directions: Firstly, a
parameterized debugging framework is introduced, which integrates
dynamic and static properties, as well as design information of
programs. The framework is based on results derived in the field of
abstract interpretation~\cite{cc77}, and can therefore be
parameterized with different lattices and context selection
strategies.

Secondly, the one--to--one correspondence between model components and
program statements is replaced by a hierarchy of components, which
provides means for more efficient reasoning procedures, as well as
more flexibility when focusing on interesting parts of a program.

This work is organized as follows. In Section~\ref{sec:mbd}, we give
an introduction to model-based debugging. Section~\ref{sec:semantics}
describes mapping from source code to model components and the
(approximate) computation of program effects in the style
of~\cite{cc77} and~\cite{bur93}. The next section discusses the
modeling of programs and the reasoning framework.  In
Section~\ref{sec:example}, we provide an example which puts together
the different models and demonstrates the debugging capabilities of
our approach.  Section~\ref{sec:impl} provides details about our
implementation.  Finally, we discuss related work and conclude the
paper.

\section{Model--based Debugging} \label{sec:mbd}

To locate faults using model--based reasoning techniques, the source
code of the program $P$ to be analyzed must be available. Also, a set
of test cases $\mathcal{TC}$ is required, which (partially) specify
the expected behavior of $P$. Test cases can be as simple as a set of
input-output vectors or even just a list of correct and incorrect
output variables.  The connection to the model--based diagnostic
framework is realized through a set $COMPS$ and a set~$\mathcal{M}$ of
models.  $COMPS$ contains the set of components of which fault
candidates are composed, whereas each $m\in\mathcal{M}$ describes the
program behavior (possibly at an abstract level) and is used to detect
discrepancies between the expected and the obtained behavior of the
program. A fault candidate in is a part of $P$'s source code that,
when assumed to show arbitrary effects, does not conflict with any
test case in $\mathcal{TC}$ any more. A fault candidate conflicts with
a test case $t$ if the modified program corresponding to the fault
candidate derives values different from the ones specifies in $t$. A
model $m\in\mathcal{M}$ of $P$ is a (partial) description of the $P$'s
behavior, derived automatically from the source code of $P$.

For example, using a model that describes dependencies between
components, where each component corresponds to a statement, the
(faulty) program
\begin{programlist}
  \plline{1}\>\keyw{int} r = 3; \\
  \plline{2}\>\keyw{float} area = r*3.141f; \\
  \plline{3}\>\keyw{float} circ = 2.f*r*3.141f;
\end{programlist} can be described as follows. 
\begin{quote}
  If statement~1 is correct, the value of \textsf{r} is correct. If
  statement~2 and \textsf{r} are correct, \textsf{area} is correct,
  too. \textsf{circ} is correct provided statement~3 and \textsf{r}
  are correct.
\end{quote}
 Translated to first order logic, this can be represented as follows:
\begin{alignat*}{1}
&\left(\neg ab(c_1)\rightarrow\mathit{correct}(r)\right)\wedge \\
&\left(\neg
  ab(c_2)\wedge\mathit{correct}(r)\rightarrow\mathit{correct}(area)\right)\wedge\\
&\left(\neg
  ab(c_3)\wedge\mathit{correct}(r)\rightarrow\mathit{correct}(circ)\right).
\end{alignat*}
$c_1$ to $c_3$ represent the components corresponding to the
statements in lines~1 to~3, respectively, and $\mathit{correct}$ is a
predicate that asserts that the variable passed as argument has the
correct value (specific to the test case under consideration). Test
cases are represented as conjunctions of $\mathit{correct}$ literals.
For example, $\mathit{correct}(circ)\wedge\neg\mathit{correct}(area)$
expresses that after running the program, variable \textsf{circ} is
correct, whereas \textsf{area} is incorrect. $ab$ is used by the
diagnostic engine to disable the model of certain components and check
if the remaining model is still inconsistent with the test case. A
more formal elaboration can be found below.

We recall some of the basic definitions from model--based
diagnosis~\cite{rei87}, slightly adapted for our purposes:
\begin{definition}[Diagnosis Problem]\label{def:diag}
  A diagnosis problem is a triple $(SD,COMPS,OBS)$ where $SD$ is the
  system description, $COMPS$ is the set of components in $SD$, and
  $OBS$ is the set of observations.
\end{definition} Here, $SD\in\mathcal{M}$ is a model of $P$, and
$OBS\in\mathcal{TC}$ is the information specified by test cases. Note
that $OBS$ contains the \emph{expected} result of test cases, not the
actual result obtained from the faulty program. Also, $OBS$ is not
restricted to pure input and output specifications; intermediate
results can also be checked using assertions (see
Section~\ref{sec:ai}).

\begin{definition}[Diagnosis]
  A set $\Delta\subseteq COMPS$ is a diagnosis for a diagnosis problem
  $(SD,COMPS,OBS)$ iff $SD\cup OBS\cup\{\neg ab(C)|C\in
  COMPS\setminus\Delta\}$ is consistent, where $ab(C)$ denotes that
  component~$C$ is not working as specified in~$SD$.
\end{definition} Each component $C\in COMPS$ corresponds to a part
of~$P$ and therefore, components in $\Delta$ indicate possible faults
in the program. The $\neg ab(C)$ behavior of a component $C$ is an
abstraction~\cite{cc77} of the semantics of the code fragment
represented by $C$, as given by the language specification. The
$ab(C)$ behavior denotes a possible fault and generally permits
arbitrary effects.

Diagnoses can be computed efficiently using the concept of conflicts,
which are sets of components that cannot be all functioning correctly
without contradicting at least one $t\in\mathcal{TC}$.
\begin{definition}[Conflict]
  $\Delta\subseteq COMPS$ is a conflict for $(SD,COMPS,OBS)$ iff
  $SD\cup OBS\cup\{\neg ab(C)|C\in\Delta\}$ is inconsistent.
\end{definition}

The basic principle of MBD is to use $\mathcal{M}$ to derive conflicts
given $\mathcal{TC}$ as observations. The conflicts are then used by
the diagnostic engine to compute diagnoses, which are mapped back to
the program's source code to indicate possible faults. To minimize the
number of fault candidates, we are only interested in subset--minimal
diagnoses, which can be derived from subset--minimal
conflicts~\cite{rei87}.

Revisiting the previous example, it is easy to see that $\{ab(c_3)\}$
cannot be a diagnosis, as the model derives the conflict $\{\neg
ab(c_1), \neg ab(c_2)\}$. However, $\{ab(c_1)\}$ and $\{ab(c_2)\}$ are
both diagnoses. $\{ab(c_1)\wedge ab(c_3)\}$ is also a diagnosis, but
not subset--minimal, as it contains the diagnosis $\{ab(c_1)\}$.

As a possible extension not covered in this paper, the approach could
be extended to output the most likely diagnoses, given prior
probabilities for each component. These probabilities can be obtained
by counting the number of correct and faulty test cases that the
statements corresponding to each component are executed
in~\cite{tip95,msww02b,jhs02}.

\section{Modeling Program Behavior} \label{sec:semantics}
  
A key aspect of every MBD system is the construction of the set
$\mathcal{M}$ of models and the mapping between the program and
$\mathcal{M}$s components. 

Previous work~\cite{msw00b,msww02a} derives the models from the source
code without considering runtime information, which often results in
large and complex models.  We construct the models dynamically, which,
by exploiting runtime information, can lead to smaller and more
concise models.

Another limitation imposed by these earlier modeling approaches is the
representation of every statement and (sub--)expression in $P$ as a
separate component in the model. Even though these models
allow for very detailed reasoning, this is rarely required in practice
and leads to a large number of diagnoses and to increased
computational requirements.

To overcome these limitations, we employ an iterative, hierarchical
diagnostic process, where the mapping from $P$ to $COMPS$ is refined
incrementally (starting with a single component for each method),
depending on the results of previous diagnostic analysis
(see~\cite{fel00b} for a similar approach). 

Previous models behave poorly when the number of loop iterations or
recursion depth is not known in advance. The combination of static
program analysis and dynamic execution proposed in the next sections
provides an effective combination, which is well--suited for dealing
with such constructs.

\subsection{Approximate Program Analysis} \label{sec:ai}

Static program analysis, in particular Abstract
Interpretation~\cite{cc77}, has successfully been applied to derive
properties of programs, even in the absence of specific test cases.
Also, the framework is customizable with different abstractions of the
concrete semantics of a program.

We recall the basic definitions of Abstract Interpretation, as given
in~\cite{cc77,bur93}:

The mapping from the concrete semantics, represented as a lattice
$(\mathcal{P}(S),\emptyset,S,\subseteq,\cup,\cap)$ ($S$ denotes the
set of program states), to the abstract, finitely represented lattice,
$(\mathcal{P^\#}(S),\bot,\top,\sqsubseteq,\sqcup,\sqcap)$, is given by
a Galois Connection $(\alpha,\gamma)$, where $\alpha$ maps sets of
states to their best approximation, and $\gamma$ maps every abstract
property to its meaning in $\mathcal{P}(S)$.

The approximate semantics of a program $P$ can then be expressed as
fixpoint over a set~$\mathcal{X}$ of equations derived from $P$'s
source code.  The equations are composed of abstract operations
$\Phi^\#\equiv\gamma\circ\Phi\circ\alpha$, which model the effects of
every operation $\Phi$ in $P$. An approximation of the forward
semantics is given by the solution of $\lfp\lambda
X\cdot(E\sqcap\mathcal{X}(X))$ (starting at $\bot$), where $E$ denotes
the approximation of the entry states. In case the abstract lattice is
of infinite height, narrowing and widening operators have to be
applied to ensure termination of the computation.  For a more
in--depth discussion see~\cite{cc77}. Bourdoncle described similar
approximations of backward semantics and added intermittent and
invariant assertions for program analysis~\cite{bur93}.

To incorporate intermittent (``sometime'') and invariant (``always'')
assertions into the analysis, a sequence of forward and backward
reasoning steps can be defined to approximate the entry and exit
states which guarantee the validity of the assertions~\cite{bur93}.
Intermittent assertions express conditions that must eventually hold
in each program execution, but not necessarily each time the program
point is reached. Invariant assertions on the other hand have to be
true every time the corresponding program point is reached (if it is
reached at all). For example, the assertion \textsf{sometime true;} at
the end of a program asserts that the program must eventually
terminate. Similarly, \textsf{always i>=0 \&\& i<=10} asserts that
whenever that point of the program is reached, \textsf{i} must be
between 0 and 10. Note that \emph{sometime} and \emph{always}
assertions, contrary to what the names may imply, do not require the
presence of multiple test cases. For example, consider a test case
where a loop executes multiple iterations. In this case, the
difference between \textsf{sometime C} and \textsf{always C} is
evident: always requires condition \textsf{C} to be true in
\emph{every} iteration, whereas sometime only requires that for
\emph{one} iteration.

\subsection{Avoiding Imprecision}

The approximation of complex programs leads to possible imprecision,
which is undesirable for automatic debugging. In particular, (1)
aliasing between variables has to be approximated, (2) it can be
difficult to derive useful properties for arrays, and (3) partitioning
of the domain of the abstraction function severely impacts the outcome
of the analysis. Further imprecision may be introduced by composition
of the abstractions for each statement. To deal with (1) and (2),
numerous different abstractions~\cite{hp00} and partial evaluation
approaches~\cite{col97} have been developed.  However, they are
generally not very well--suited for MBD, because the results are often
too imprecise to derive a conflict. To overcome (3),
\cite{bur92}~introduced a model that is able to refine the domain
based on the current partitioning. However, even in this framework,
the choice of approximation operators remains crucial (and program
dependent).

To circumvent the aforementioned shortcomings, we employ the
information from test cases to avoid approximation whenever possible,
and rely on static analysis only as a fallback in case the program's
behavior is only partially specified or exceeds user--defined bounds.
A more detailed discussion is provided in Section~\ref{sec:modeling}.

\section{Model Construction} \label{sec:modeling}

In this section, we present a model that follows the execution
semantics of the program. Based on the semantic approximation
introduced in the previous section, a separate model for each test
case is constructed by abstract interpretation of the program, using
the entry state specified by the test case. Test case information
(pre- and postconditions values, as well as intermediate assertions)
is mapped to \emph{sometime} and \emph{always} assertions. This
differs from traditional abstract interpretation
techniques~\cite{cc77} as we generate the equations representing the
system dynamically while the fixpoint is computed, which is
advantageous when combined with the MBD engine and partitioning
strategies (see below). The model derives a contradiction iff there
exists no feasible path between the entry state and the exit state of
the program.\footnote{Here, the assumption is made that the program is
  terminating; this issue is revisited below.} To determine the set of
components the conflict is composed of, we follow the approach
of~\cite{mt02a}. The algorithm can be summarized as follows. After a
conflict has been detected, the derivation tree is analyzed to find
the subset--minimal set of constraint needed to derive the
inconsistency. This is done by recursively subdividing the derivation
tree and pruning sets that cannot contribute to the conflict.

The dynamic approach, together with the test case information allows
us to explore only these parts of the model which may actually be
executed.  Especially for object--oriented languages like Java, with
many possibly exception--throwing statements, this approach results in
significantly smaller models. For example, if a branch of a
conditional can be eliminated, its statement need not be considered
and the data flow $\phi$ and $\sigma$ functions~\cite{ananian99} can
be eliminated, too.

\subsection{Partitioning Strategies}
\label{sec:partitionstrategies}

Crucial to the accuracy of the results is the selection of
partitioning strategies for contexts of method calls. This corresponds
to the selection of widening operators in~\cite{bur92}.  We propose a
heuristic strategy that introduces a new partition whenever the call
is non--recursive or the calling statement is definitely executed for
every possible execution of the model, and a common partition
representing the called method otherwise. The strategy can be further
enhanced by bounding the depth of the call stack, possibly with
different bounds for different categories of methods. The analysis and
identification of useful heuristics constitutes an important part of
future model refinement. To keep the analysis feasible, sparse
representation of environments have to be used (see
Section~\ref{sec:impl} for more details).

Another key feature in the analysis of object--oriented programs is
the abstraction of heap data structures and aliased variables.  For
abstracting heap data structures, any of the numerous heap abstraction
approaches developed in the last decades can be applied.  For
simplicity, we propose the approach given by~\cite{cor98}, where
objects are abstracted into equivalence classes associated with the
program point at which they were created. Note that simple approaches
can lead to accurate results, as the partitioning strategies, together
with the information from test cases, in many cases eliminate the need
for approximation.

\subsection{Analyzing Loops}

For simplicity, we restrict the following discussion to \textsf{while}
loops (other forms of iteration statements are treated similarly).

In case the condition of the loop can be evaluated uniquely using the
abstract environment, the corresponding branch is followed, unless a
termination check is triggered. Otherwise, conventional static
analysis of the loop is done.  The environment before and after the
loop, together with assertions from the test case, are used as entry
and exit states for the analysis.  Static analysis is used to
strengthen the pre- and postconditions of the loop, which are
subsequently used to derive conflicts.

Note that in this approach, more precise results than with static
analysis alone can be derived. This is because the values
in the pre- and post environment are derived from test cases and,
therefore, may be more precise than a purely static approximation.
These values can in turn be used to derive approximation operators for
the static analysis, which are effective at proving a contradiction.
This idea is similar to~\cite{cdhprlz00}, where abstraction operators
are guessed based on the structure of the program and the given proof
obligation.

\label{sec:termination}

Nontermination,\footnote{Although we assume the given
  program is terminating on all test cases, nonterminating programs
  can arise due to abnormality assumptions set by the diagnostic
  engine.} by necessity, has to be dealt with through heuristics,
such as setting upper limits on loop iterations or recursion depth, or
using the abstract interpretation model to determine if any of the
successor statements of the loop (call) can be reached using the
current environment as entry state for the loop (call). This is an
avenue for future research.

\subsection{Correcting Faults}
\label{sec:replacements}

Once the possible fault locations have been narrowed down to a single
candidate, heuristic algorithms to guess a replacement for the faulty
instruction are applied. From the values of the environment before and
after the faulty statement, and from the statements internal
structure, instructions are synthesized to replace the incorrect
statement~\cite{sw99a}. This process can be aided by complementary
models (Section~\ref{sec:complmodels}), to restrict the search space
for candidate instructions. We briefly sketch the
synthesis algorithm; see~\cite{sw99a} for further details. 

Let $s$ be the statement to be replaced. The set of replacement
statements for $s$ is derived according to the type of $s$ and is
constrained by a parameter~$k$ that limits the maximum size of the
replacement. Replacements that do not satisfy the static type
declarations in the original program are not considered, as we are
only interested in corrections that satisfy the basic requirements of
the language and the compiler. Possible replacements for constants and
variables are constants, different variables, or calls to methods
defined in the program. Method calls can be replaced either by a call
to a different method (using a subset of the original calls arguments,
or synthesizing new arguments), or by one of it's arguments (if any).
For each replacement, a penalty measure (``size'') is assigned, which
measures the deviation from the original program, with zero being no
modification.

The algorithm finds suitable replacements by enumerating possible
replacements up to size~$k$, ignoring all candidates that are
inconsistent with the types and values derived for the diagnosis
candidate $ab(s)$.  Thus replacements more similar to the original
program are tested earlier and are preferred to less similar
statements. This algorithm can be extended to incorporate information
provided by complementary models, as is demonstrated in the example in
Section~\ref{sec:example}. In this case, variables that are indicated
by the complementary model are assigned lower penalty values than
other variables, resulting in former candidates being preferred.

For example, statement \mbox{\textsf{\keyw{float} circ =
    2.f*r*3.141f}} has the following replacement candidates: replacing
either constant with another constant (size~1) or with a variable
(size~2), or with a method call without arguments (size~2).
\textsf{r} can be replaced by a constant, another variable, or a
method call without arguments (size~1, 1, and 2, respectively).
Operators can be replaced with any of their arguments, a different
operator, or a method call taking two arguments.  Finally, the
variable on the left hand side of the assignment can be replaced with
any other assignable variable of the same type.

The modification of the model to incorporate the replacement
instructions is done by introducing specialized mode assumptions
\syn{$l$}{$r$} and \nosyn{$l$}, where $l$ is the program point where
the replacement $r$ is applied. \nosyn{$l$} expressed that no
modification takes place at $l$.

\subsection{Complementary Models} \label{sec:complmodels}
Past experiences with MBD have shown that MBD provides excellent
results when diagnosing functional faults. However, for structural
faults, the semantics--based models do not provide sufficient
information to accurately detect such faults. Furthermore, unless the
test case specification is unreasonably detailed, for many programs a
large number of diagnoses remains.

To overcome these problems, \cite{mst01a}~proposed to utilize
complementary models, in particular representations of design
information, to obtain the necessary information to guide the
diagnostic engine. An advantage of this approach is that it integrates
nicely into our framework without placing additional burden on the
user.

For example, consider the state diagram in
Figure~\ref{fig:sequencediag}. The automaton can be interpreted as a
specification expressing that for each object the first call to method
\textsf{getValue} (if any) must be preceded by a call to
\textsf{setValue}. Translated into assertion statements and
incorporated into our debugging engine, this model can be used to
detect wrong or missing method calls, as demonstrated in the next
section.

Specifically, we propose to use partial specifications, i.e.\ pre- and
postconditions and assertions, to generate additional conflicts. This
provides multiple advantages:
\begin{itemize}
\item Paths of the model can be eliminated and possibly new conflicts
  be generated by checking the assertions.
\item Assertions can be valid for multiple test cases, which avoids
  specifying program behavior separately for every test case and
  values. 
\item By comparing the dependencies between variables in pre- and
  postconditions, structural faults, such as wrong assignments or
  missing statements, can be detected~\cite{mst01a}. A similar
  approach, but without exploiting test case information, was used
  in~\cite{jackson95}.
\end{itemize}

\section{Example}
\label{sec:example}

\begin{figure}
  \begin{center}
    \begin{minipage}{.5\textwidth}
      \begin{programlist}
        \plline{1}\>\keyw{class} Item \{ \\
        \plline{2}\>\>\keyw{int} value;   \\
        \plline{3}\>\>\keyw{void} setValue(\keyw{int} v) \{ value = v; \} \\
        \plline{4}\>\>\keyw{int} getValue() \{ \keyw{return} value; \}     \\
        \plline{5}\>\} \\
        \plline{6}\>\keyw{class} Main \{ \\
        \plline{7}\>\>Item[] items; \\
        \plline{8}\>\>\keyw{int} first, last; \\
        \plline{9}\>\>/**  @pre: (n \ensuremath{>} 0) */ \\
        \plline{10}\>\>\keyw{void} setup(\keyw{int} n, \keyw{int} d) \{ \\
        \plline{11}\>\>\>\keyw{int} i = 0; \\
        \plline{12}\>\>\>\keyw{int} k = 1; \\
        \plline{13}\>\>\>items = \keyw{new} Item[n]; \\
        \plline{14}\>\>\>\keyw{while} (i \ensuremath{<} items.length) \{ \\
        \plline{15}\>\>\>\>Item item = \keyw{new} Item(); \\
        \plline{16}\>\>\>\>items[i] = item; \\
        \plline{17}\>\>\>\>k *= d; \\
        \plline{18}\>\>\>\>i++; \\
        \plline{19}\>\>\>\} \\
        \plline{20}\>\>\>first = items[0].getValue(); \\
        \plline{21}\>\>\>last = items[n--1].getValue(); \\
        \plline{22}\>\>\} \\
        \plline{23}\>\>/** @post: (first == d) \&\&  \\
        \plline{24}\>\>\ *   (last == Math.pow(d,n)) \&\&  \\
        \plline{25}\>\>\ *   (items.length == n) */ \\
        \plline{26}\>\} \\
      \end{programlist}
    \end{minipage}
  \end{center}
  \caption{Example Program}
  \label{fig:program}
\end{figure}

This section puts together all the previous sections and demonstrates
the framework's ability to locate and correct faults.

In this example, the interval abstraction from~\cite{cc77} is used to
approximate a set of integer values. The model is structured such that
diagnosis components represent single statements. For simplicity,
hierarchic modeling is not applied, as the method's structure is
rather simple. Also, the example does not require termination
heuristics for loops or method calls.  For objects created on the
heap, a simple abstraction aggregating all objects created at a
specific location into one abstract model variable, corresponding to
the allocation site, is used~\cite{cor98}.

\begin{figure}
  \begin{center}
    \includegraphics[scale=.6]{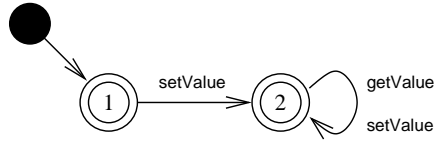}
  \end{center} \caption{Valid Call Sequences for \textsf{Item}}
  \label{fig:sequencediag}
\end{figure}

Consider the program in Figure~\ref{fig:program}, where the statement
\textsf{item.setValue(k)} is missing after line~17. Further, assume a
test case $T=\{\stmt{10}\mapsto\langle
n=3,d=2\rangle,\stmt{22}\mapsto\langle items[k].value=2^{k+1}
(k\in[0..2])\rangle\}$\footnote{Components corresponding to statements
  are identified by the statement's line number.} and the contract
specification given in the method comments.  Note that the test case
\emph{specification} expresses the \emph{intended} result of the
program. Also, we are given a complementary model that specifies that
for each instance of class \textsf{Item}, the method
\textsf{setValue()} must be called before \textsf{getValue()} is
invoked (see Figure~\ref{fig:sequencediag}). This is translated into a
separate instance variable \textsf{Item.\_callstate}, which is
initialized to~1, denoting the state before the first method call in
Figure~\ref{fig:sequencediag}. At the entry point of method
\textsf{setValue()} \textsf{\_callstate=2} is inserted, indicating that
after the method is called, the automaton in
Figure~\ref{fig:sequencediag} is in state~2. Similarly, in method
\textsf{getValue()}, the assertion \textsf{always \_callstate==2} is
inserted.\footnote{Note that the mapping from sequence diagrams to
  assertions is more complicated if a state diagram contains multiple
  paths leading to a method. For simplicity, we refrain from a
  detailed discussion in this paper.} 

When the method \textsf{setup()} is analyzed using the test case, the
conflict $\mathcal{C}=\{\nab{13},\nab{15},$ $\nab{20}\}$ is derived:
\stmt{12}, \stmt{17}, and \stmt{3} do not influence the result or the
call sequence at all and can therefore be removed; when \stmt{4},
\stmt{11}, \stmt{14}, \stmt{16}, and~\stmt{18} are abnormal, the
complementary model in Figure~\ref{fig:sequencediag} still derives a
conflict with~\stmt{20}, as \textsf{setValue()} is not called before
\textsf{getValue()}; \stmt{21} can be removed for the same reason.

The diagnostic process continues with the assumption that at least one
component in $\mathcal{C}$ is faulty. Rerunning the model for each
component, the following conflicts are derived: \ab{13} conflicts with
\{\nab{4},\nab{11},\nab{14},\nab{15},\nab{16},\nab{18},\nab{20}\}
because replacing \textsf{\keyw{new}~Item[n]} with another expression
still causes a contradiction for \textsf{first} in line 20, or a
\textsf{NullPointerException}. For \ab{15}, no type compatible
replacement for \textsf{\keyw{new} Item()} exists and, therefore, this
assumption is not consistent either.  \ab{20} induces the conflict
\{\nab{13},\nab{15},\nab{16},\nab{21}\}, as \stmt{21} either causes a
\textsf{NullPointerException}, or the the complementary model again
derives a call sequence conflict.

As none of the attempts to restore consistency by assuming abnormality
for any of the components of the initial conflict is successful, no
single--fault diagnosis exists for the given program and test case. As
a consequence, the diagnostic process has to choose between increasing
the diagnosis cardinality or to search for structural faults. As we
are interested in simple faults, it is reasonable to look for
structural faults before increasing diagnosis cardinality.

Using the model from Figure~\ref{fig:sequencediag}, and from the
conflicts above, it can be deduced that a call to \textsf{setValue()}
is missing. From the conflict $\mathcal{C}$ and the knowledge that the
objects causing a contradiction were created in \stmt{15}, the missing
call can be inserted between~\stmt{15} and~\stmt{20} (denoted
\stmt{$15'$}---\stmt{$19'$}).

The diagnostic process is restarted, using four new rules to insert a
synthesized statement $s'$ (Section~\ref{sec:replacements}) at
location $l$ whenever \syn{$l$}{$s'$} is assumed (with \nosyn{$l$}
being the default).  The simplest candidates including a call to
\textsf{setValue()} are of the form
\textsf{$\alpha$.setValue($\beta$)}, where
$\alpha\in$\{\textsf{item},\textsf{items[$\alpha'$]}\},
$\alpha',\beta\in$\{\textsf{first},\textsf{last},\textsf{i},\textsf{k},\textsf{n},
\textsf{d}\}.

To further restrict the synthesis candidates, we utilize dependency
information provided by a complementary model: The postconditions of
the method imply a dependency from variable \textsf{d} to the
variables \textsf{first} and \textsf{last}. On the other hand, these
dependencies cannot be derived from the implementation. As
\textsf{first} and \textsf{last} depend on
\textsf{items[$\cdot$].value} only, either \textsf{first} and
\textsf{last} directly, or \textsf{items[$\cdot$].value} must also
depend on~\textsf{d}. Therefore, the synthesized statements using
\textsf{d} or \textsf{k} as argument to \textsf{setValue()} are
preferred. This example also illustrates  that it can be
advantageous to express assertions about programs in terms of
variables of the program instead of test case specific values, where
comparing dependencies is not possible.

With models that have been modified by adding the synthesized
expressions, three diagnoses are obtained:
\syn{$15'$}{\textsf{items[i].setValue(k)}} and
\syn{$l$}{\textsf{item.setValue(k)}} with $l\in\{17',18'\}$. Other
candidates are not consistent for the following reasons:

\begin{tabular}{lll}
Location & Candidate & Conflict \\ \hline
\stmt{$15'$} --- \stmt{$18'$} & \textsf{item.setValue(d)} &  Contradiction with test case (for \textsf{first} or \textsf{last}) \\
\stmt{$15'$}, \stmt{$16'$} & \textsf{item.setValue(k)} &  Contradiction with test case (for \textsf{first} or \textsf{last}) \\
\stmt{$17'$}, \stmt{$18'$} & \textsf{item.setValue(k)} &  --- \\
\stmt{$15'$} & \textsf{items[$\alpha$].setValue($\beta$)} & Uncaught \textsf{NullPointerException} \\
\stmt{$16'$} --- \stmt{$19'$} & \textsf{items[$\gamma$].setValue($\beta$)} &  Contradiction with test case (for \textsf{first} or \textsf{last}) \\
\stmt{$16'$} --- \stmt{$19'$} & \textsf{items[k].setValue($\beta$)} &  Uncaught \textsf{ArrayIndexOutOfBoundsException} \\
\stmt{$18'$}, \stmt{$19'$} & \textsf{items[i].setValue($\beta$)} &  Uncaught \textsf{ArrayIndexOutOfBoundsException} \\
\stmt{$16'$} & \textsf{items[i].setValue($\beta$)} &  Contradiction with test case (for \textsf{first} or \textsf{last}) \\
\stmt{$17'$} & \textsf{items[i].setValue(d)} &  Contradiction with test case (for \textsf{first} or \textsf{last}) \\
\stmt{$17'$} & \textsf{items[i].setValue(k)} &  --- \\
& \multicolumn{2}{l}{($\alpha\in\{$\textsf{d,k,n,first,last}$\}$, $\beta\in\{$\textsf{d},\textsf{k}$\}$,$\gamma\in\{$\textsf{d,n,first,last}$\}$)} \\
\end{tabular}

Note that the combination of dynamic execution and static analysis is
more powerful than static analysis alone. As a demonstration, 
consider the synthesized statement \textsf{items[d].setValue(d)}.
Our model is able to derive a conflict with variable \textsf{first},
whereas static analysis cannot because elements of the array are
approximated as $[0..\mathsf{d}]$ (assuming an interval
abstraction~\cite{cc77} for \textsf{Item.value}, and a heap
abstraction that does not distinguish between \textsf{items[0]} and
\textsf{items[d]}).

Finally, the suggestion to insert either \textsf{items[i].setValue(k)}
after line~17, or \textsf{item.setValue(k)} after lines~17 or~18 is
presented to the user.

\section{Sparse Trace Representation}
\label{sec:impl}

This section describes a generalized notion of program trace and a
sparse representation thereof. This representation makes it possible
to avoid copying parts of the dynamic data structures created by a
program, as was required by previous models~\cite{mayer00}.

\begin{definition}[Variable Identifier]
  A variable identifier is either the canonical name of a local or
  static variable, or is composed of an object identifier~$o$ and a
  canonical name of an instance variable~$v$ (denoted~$o.v$).
\end{definition}

The canonical name of a variable is formed by prefixing its name with
the fully qualified name of the scope the variable is defined in. For
example, the canonical name of static variable \textsf{out} defined in
class \textsf{System}, which is defined in package \textsf{java.lang},
is \textsf{java.lang.System.out}. Object identifiers are an
abstraction of memory addresses for objects created on the heap. In
this work, we use the statement that created the object as
identifier. For multi--dimensional arrays, the index of the parent
array is included to distinguish different sub--arrays.

\begin{definition}[Environment]
  An environment $e$ is a tuple $\langle c:l,\mathcal{V}\rangle$,
  where $c\in\mathcal{C}$ is a unique context identifier and
  $l\in\mathcal{L}$ the label of the statement $e$ is associated with.
  $\mathcal{V}$ is a mapping from variable identifiers into abstract
  values.  $\mathcal{E}$ denotes the set of all abstract
  environments~$\mathcal{V}$.
\end{definition}

In this work, context and partition are used in the sense of points-to
analysis or call graph construction~\cite{gfdc97} and represent an
abstraction of the call site of a method. Context identifiers are used
to distinguish different instances of a program part during analysis.
For example, if a method is called multiple times in a trace, the
analysis of the two calls can be merged into a single analysis, using
the merged input and output environments of both calls. While speeding
up the analysis, merging, i.e.  partitioning, contexts results in
diminished accuracy and is applied only when necessary. Traces without
loops can be analyzed without merging, while recursive calls or loop
statements may require approximation in case the recursion depth or
the number of iterations cannot be determined.

The relationship between concrete and abstract values is given by the
abstraction function selected for the abstract interpretation. An
abstract environment associates each variable identifier with an
abstract value, thus approximating the set of concrete values in a
non--relational way. For example, the well--known interval
abstraction~\cite{cc77} approximates a set of integer values with an
interval spanning all the values in the set.

Every program is transformed into a simple intermediate
representation, consisting of assignments, primitive operations and
method calls at top--level. 
\begin{definition}[Program]
  A program $P$ is a pair $\langle\mathcal{S},\mathcal{R}\rangle$,
  where $\mathcal{S}$ is a finite set of statements $l:s$, each
  labeled with a unique label $l\in\mathcal{L}$.
  $\mathcal{R}\subseteq\Gamma(\mathcal{E})\times\Gamma(\mathcal{E})$
  is a transfer relation, specifying the possible transitions between
  concrete environments. $\Gamma(\mathcal{E})\stackrel{\mbox{\tiny
    def}}{=}\{\gamma(e) | e\in\mathcal{E}\}$.
\end{definition} For short, $\langle a,b\rangle\in\mathcal{R}$ is
denoted $a\stackrel{\mathcal{R}}{\rightarrow} b$.

A context selection function $C$ generates a label for the destination
environment, given an environment.

\begin{definition}[Execution Trace]\label{def:trace}
  The execution trace $\mathcal{T}$ of $P$ is defined as
  $\mathcal{T}=\bigcup_{i\ge 0}\mathcal{T}^i$, with
  $\mathcal{T}^0=\{\exists_p
  \gamma(e_0)\stackrel{\mathcal{R}}{\rightarrow}p\}$,
  $\mathcal{T}^{i+1}=\mathcal{T}^i\cup\{p\rightarrow q|\exists_r
  r\stackrel{\mathcal{T}^i}{\rightarrow}
  p,\gamma(p)\stackrel{\mathcal{R}}{\rightarrow}\gamma(s),s=\langle
  c:l,\mathcal{V}\rangle, c'=C(s), q=\langle
  c':l,\mathit{map}(\mathcal{T}^i,c':l)\sqcup\mathcal{V}\rangle\}$.
  $e_0$ denotes the abstract environment at the starting point of the
  trace.  $\mathit{map}(\mathcal{T}^i,c':l)$ denotes the variable
  mapping for the environment labeled $c':l$ in $\mathcal{T}^i$
  ($\bot$ if none exists), and $\sqcup$ is the join operator of the
  abstract domain lattice.
\end{definition} 
This definition builds the graph containing all feasible paths,
starting from $e_0$. Given the reachable environments from the
previous iteration, new transitions are added leading to the reachable
environments as specified by $\mathcal{R}$. 

The context selection function $C:\mathcal{E}\mapsto\mathcal{C}$
determines the context of the target environment, given the source
environment. For recursive method calls and loops, infinite execution
sequences have to be finitely approximated by partitioning the set of
all execution contexts into finitely many partitions. $C$ can be
influenced by the user or by heuristics to adjust the degree of
imprecision.  
As mentioned in Section~\ref{sec:termination}, a bounded approximation
of the call stack and loop counter can be used to analyze recursive
and iterative program constructs. For other program elements, $C=Id$
is sufficient, as no approximation is necessary.

Building on the (\emph{Static Single Information} form
(SSI)~\cite{ananian99,ms02a}, a separate copy of all used variables is
created for each branch in the execution trace.  The SSI form makes
use of $\phi$ and $\sigma$ functions, which deal with data flow
information spanning control flow paths. $\phi$ functions are placed
at control flow join points and combine the values from the incoming
branches into an approximate value that is used for references to that
variable below the $\phi$ function. Similarly, $\sigma$ functions are
placed at control flow branches and create separate copies of the
incoming value for each branch. Instead of computing locations for
$\sigma$ and $\phi$ functions statically, we utilize test case
information to restrict execution paths. This makes it possible to
omit many of the $\sigma$ and $\phi$ functions, which in turn produces
simpler models.

\subsection{Incremental Trace Construction} 

To allow for efficient computation of the used and modified variables,
the trace representation is split into two parts. One part is a static
approximation of all possible traces, used to determine basic blocks
where $\phi$ and $\sigma$ functions may be necessary. The other part
represents the feasible execution paths through the program, starting
from $e_0$.

By constructing the trace incrementally while executing a test case,
the used and modified variables, and in particular, the used and
modified objects, are known. As a consequence, only the objects that
are actually modified by a statement need to be updated, leaving all
other objects unmodified. Consequently, subsequent statements are
connected directly to the last modification of their used objects,
instead of being connected to the last modification of \emph{any}
instance of the same type. Therefore, copying values of unmodified
instances is unnecessary and can be omitted.

The static approximation $\mathcal{T}_S$ of $C\circ\mathcal{R}$ for
each method is derived from a control flow graph (CFG) as described
in~\cite{ms02a}.  This is possible as
Section~\ref{sec:partitionstrategies} restricts $C$ such that only for
method entry nodes or loop headers, $C\neq Id$ is possible. From the
approximation, the \emph{DF~graph}~\cite{sre95} is constructed. The DF
graph contains all nodes of the trace's CFG, indicating nodes which
are locations of possible $\phi$ functions. From each node $n$ in the
graph, links point to the $\phi$ functions in the dominance frontier
of $n$.\footnote{The dominance frontier of a node $n$ contains all
  nodes $n'$ of which $n$ dominates an immediate predecessor of $n'$,
  but not $n'$ itself. It can be shown that these are exactly the
  locations where $\phi$ functions need to be placed.}  Therefore, by
reversing the direction of the links, each $\phi$ function is linked
to the collection of nodes which give rise to $\phi$. Each node is
labeled with an unique identifier.

The representation $\mathcal{T}_D$ of all feasible execution paths is
built by repeatedly applying Definition~\ref{def:trace}, starting from
the entry environment $e_0$.  Transitions are grouped into basic
blocks, with linear transition sequences\footnote{A linear transition
  sequence is a nonempty sequence $\langle e_i\rightarrow
  e_{i+1}\rightarrow\dots\rightarrow e_{i+k}\rangle$, where none of
  the $e_{i+1},\dots,e_{i+k-1}$ has more than one predecessor or
  successor.} being compressed into one block.

Environments with multiple outgoing transitions in $\mathcal{T}_D$
denote the end of the current basic block. For each outgoing
transition $p\rightarrow q$ with $q\not\in\mathcal{T}_D$, new basic
blocks are created. Otherwise, $q$ already exists and just gains
a new incoming link. At this point, several steps are necessary to
preserve the correctness of $\mathcal{T}_D$:

First, in case $q$ is not the first environment of a block $B$, $B$ is
split into two parts, $B_0$ and $B_1$, consisting of the transitions
leading to $q$ and the remaining path starting at $q$, respectively.
$B_1$ is inserted as a successor of $B_0$ and the link $p\rightarrow
q$ is added.

Next, the set $\mathcal{B}=DF^{-1}(q)$ of blocks giving rise to a
$\phi$ function at this environment $q$ is determined using the links
in $\mathcal{T}_S$, where only blocks are considered that are actually
instantiated in $\mathcal{T}_D$ for the current context. For each
$b\in\mathcal{B}$, the set of modified variables is determined and
$\phi$ functions are created for each variable (unless they already
exist).

To maintain correctness of $\mathcal{T}_D$, the ordering in which the
transitions are processed is crucial. It must be ensured that all
modified variables for a block are known before the block is used to
generate other $\phi$ functions. This can be ensured by suspending the
processing of $\phi$ function generation, in case not all blocks of
the current context corresponding to blocks in $\mathcal{B}$ have been
analyzed completely, or are known to be unreachable. In addition,
an ordering has to be imposed on contexts and labels, such that loops
and called methods are analyzed completely before any of the successor
transitions are expanded. If assuming the proposed context selection
strategy from Section~\ref{sec:partitionstrategies}, this is not a
severe restriction for our framework.

If the graph is cyclic, this ordering is not enough, and a
fixpoint algorithm needs to be used (details are omitted for brevity).

The introduction of $\sigma$ functions~\cite{ananian99} for used
variables is handled similarly to~$\phi$ functions. Possible locations
for $\sigma$ functions are computed using $\mathcal{T}_S$ with the
direction of all arcs reversed, and an auxiliary environment that
postdominates\footnote{An environment $e_1$ postdominates $e_2$ iff
  all paths from $e_2$ to the exit environment visit $e_1$.} all exit
environments in the original $\mathcal{T}_S$.  Special treatment of
cyclic structures is not necessary in this case.

Whenever a branch of the trace is found inconsistent, the branch is
removed and replaced with a summary of the part of the derivation of
the inconsistency that is local to the branch. In case the branch is
the only outgoing or incoming connection to a $\sigma$ or $\phi$
function, the function is removed and all used associated variable are
redirected to the previous definition. Although branch elimination is
not necessary for the initial forward trace construction (as only
consistent branches are followed), existing inconsistent branches may
be found in subsequent backward and forward iterations.

\subsection{Complexity}

The time complexity of the trace construction is
$\mathcal{O}(n_D\cdot\max(n_S,n_D)^2\cdot\alpha(n_D))$ in the worst
case, where $n_S$ and $n_D$ denote the number of blocks in
$\mathcal{T}_S$ and the number of environments in $\mathcal{T}_D$,
respectively.  $\alpha(n_D)$ represents the worst case complexity of
the fixpoint computation, which depends on the program structure and
on the abstract domain lattice.

\section{Related Work}
\label{sec:relwork}

Automated debugging has been an active area of research for several
decades, resulting in a large number of different methodologies using
various assumptions and algorithms.

In Program Slicing~\cite{wei84,tip95}, statements that cannot
influence the value of a variable at a given program point are
eliminated by considering the dependencies between the statements.
Backward reasoning from output values, as in our approach, is not
possible. Similar ideas were successfully utilized in a MBD tool
analyzing VHDL programs~\cite{fsw99,wot01b}.

\cite{burhor93,burhor95} use probability measurements to guide
diagnosis. The program debugging process is divided into two steps. In
the first one, program parts that may cause a discrepancy are computed
by tracing the incorrect output back to the inputs and collecting the
involved statements. In a second step, a belief network is used to
identify the most probable statements causing the fault. Although this
approach was successful in debugging a very large program, it requires
statistics relating the statement types and fault symptoms, which
makes it unsuitable for debugging general programs. 

The idea of path information to guide debugging was also applied by
other researchers, such as program dicing~\cite{tip95} and similar
heuristics~\cite{ps92} and visualization of test results~\cite{jhs02}.
Whereas those ideas seem to provide good results, they are even more
valuable when integrated into a model--based debugging environment, as
they can provide the necessary information to discriminate between
diagnoses and aid the selection of more likely
candidates~\cite{msww02b}.

Jackson~\cite{jackson95} introduces a framework to detect faults in
programs that manifest through changed dependencies between the input
and the output variables of a program. The approach detects
differences between the dependencies computed for a program and the
dependencies specified by the user. It is able to detect certain kinds
of structural faults but no test case information is exploited.
Whereas Jackson focuses on bug detection, the model--based approach is
also capable of locating faults. Further, the information obtained
from present and absent dependencies can aid the debugger to focus on
certain regions and types of faults, and thus find possible causes
more quickly.

\cite{hunt98} applies the idea of MBD to the domain of object--oriented
languages by building models for programs written in Smalltalk. The
model used in his work is based on dependencies between instance
variables and method calls that modify them. The observations state
whether the computed value of a variable is correct or not, regardless
of its concrete value. This approach is limited to programs that
contain a single faulty statement. Also, previous results
showed~\cite{wie01} that while dependencies are a valuable tool to
isolate faulty modules, more expressive models are needed to locate
faults on a finer--grained level and to reduce frequent
user--interaction.

\cite{hz00} introduces an algorithm that compares a faulty program to
a close correct variant to determine changes that cause the
misbehavior. Although the algorithm seems to be highly effective for
test case minimization and has also been applied to locate failure
causes in programs~\cite{cz00}, the approach generally requires a
close and correct variant of the program (or a preselection of
``interesting'' statements, for example in form of grouped changes
from a versioning system) to be effective.

\cite{con93} were the first to study model--based debugging, with logic
programs as language of interest. Their approach was later extended
and refined by~\cite{bond94b, bond94a}. Their approach connects
diagnosis and debugging by identifying horn clauses to be added or
removed from programs to fix a fault. They show that the MBD approach
is more efficient in terms of user interaction than Algorithmic
Debugging~\cite{sha83}.

\cite{fel00b} apply similar ideas to knowledge base maintenance, exploiting
hierarchical information to speed up the diagnostic
process and to reduce the number of diagnoses. 

Following \cite{con93}, MBD was extended to imperative and concurrent
languages, in particular to a subset of VHDL~\cite{fsw99}. This
work showed that MBD can be successfully applied in this domain to
isolate faulty processes. Diagnosing programs at a finer level of
granularity is still ongoing research~\cite{pw03} and requires
overcoming difficulties related to temporal and concurrency-related
aspects of the VHDL language.

Mateis~et~al.~\cite{msw00b} introduce a dependency--based model
for Java programs that abstracts from concrete variable values.
However, for programs with complex structure, either a high amount of
user--provided information is necessary, or the results are relatively
coarse. In~\cite{mayer00}
it was extended to simulate program
execution. The models are limited to structured, non--recursive
programs and are not as expressive as the
abstract--interpretation--based approach when the behavior of complex
components is only partially deducible given a test case and
diagnostic assumptions.

Previous research in MBD has resulted in a set of tools that
successfully demonstrated the potential of the approach. The main
strength of the model--based techniques is that reasoning strategies
are separated from conflict detection, which makes it feasible to
plug--in a variety of program analysis and debugging methods, provided
the results of the analysis can be mapped back to the program's source
code. A number of models have been developed and analyzed, resulting
in promising results, mainly in the domain of functional faults (such
as wrong constants, operators, conditional expressions, etc.).
However, the combination of multiple models and reasoning strategies
to improve accuracy and reduce user interaction is still ongoing
research and needs further evaluation, in particular with a larger set
of realistic programs. This work aims at making a first step in this
direction by combining abstract--interpretation--based models with
complementary models to correct omitted statements and structural
faults. Also, the implementation of most of the models currently is
only incomplete and experimental. In particular, no optimizations for
speed have been done, which makes the comparison with other approaches
rather difficult.

Abstract Interpretation to analyze programs was first introduced
by~\cite{cc77}, and later extended by~\cite{bur93,cc00} to include
assertions for abstract debugging. Their approach aims at analyzing
every possible execution of a program, which makes is suitable to
detect errors even in the case where no test cases are available. A
common problem of these approaches is that of choosing appropriate
abstractions in order to obtain useful results, which hinders the
automatic applicability of these approaches for many programs.
\cite{bur92} introduces a relaxed form of representation for abstract
interpretation, which allows for more complex domains, while building
the structure of the approximation dynamically. Our framework is
strongly inspired by this work, but provides more insight on how to
choose approximation operators for debugging, in particular in the
case where test information is known. These questions are not
addressed in~\cite{bur92}.

Recently, model checking approaches have been extended to attempt
fault localization in counterexample traces. \cite{bnr03} extended a
model checking algorithm that is able to pinpoint transitions in
traces responsible for a faulty behavior. \cite{gv03} presents another
approach, which explores the neighborhood of counterexamples to
determine causes of faulty behavior. These techniques mostly consider
deviations in control flow and do not take data dependencies into
account. Also, the derivation of the abstract model from the concrete
program usually is non--trivial and is difficult to automate.

\section{Conclusion}
\label{sec:conclusion}

We have presented an automatic debugging approach utilizing
model--based diagnosis together with an abstract interpretation based
conflict detection framework. Based on experiences with previous
models~\cite{msww02a}, this framework is able to detect large classes
of programming errors, such as faulty expressions and faults in
control flow, given a set of test cases and partial specifications of
the programs behavior. This work extends the approach to provide more
accurate results in cases where previous models could not derive
conflicts by approximating loops and recursive function calls using
abstract interpretation. Further, the introduction of complementary
models allows to extend this approach to structural faults.  The
abstract interpretation framework makes it possible to parameterize
the framework in various directions: the approximation of variable
values can be chosen, heuristics for partitioning of context for
static analysis and heap analysis are parameterizable, and heuristics
for detection of nontermination are incorporated to avoid
nonterminating diagnoses. The framework's ability to locate and
correct certain faults automatically was demonstrated using a simple
example program.  Possible extensions are the representation for
abstract domains from~\cite{bur92}, and the analysis and refinement of
heuristics for context partitioning and termination detection. While
those heuristics are not essential for our approach, abstractions
tailored to specific programs and specifications~\cite{cdhprlz00} can
improve the results dramatically.

\bibliography{references}

\end{document}